\documentclass[dvipsnames,aps,prl,nobibnotes,twocolumn,superscriptaddress,nolongbibliography]{revtex4-1}

\usepackage[T1]{fontenc}
\usepackage[english]{babel}
\usepackage[utf8]{inputenc}

\usepackage[unicode=true,pdfusetitle,bookmarks=true,bookmarksnumbered=false,bookmarksopen=false,
 breaklinks=false,pdfborder={0 0 0},pdfborderstyle={},backref=false,colorlinks=true]
 {hyperref}
 \hypersetup{
 citecolor=blue,
 urlcolor=blue,
 linkcolor=blue}

\usepackage{array}
\usepackage{wrapfig}
\usepackage{float}
\usepackage{units}
\usepackage{amsmath, relsize}
\usepackage{cancel}
\usepackage{mathrsfs}
\usepackage{amssymb}
\usepackage{graphicx}
\usepackage{stackrel}
\usepackage{xparse}
\usepackage{soul}
\usepackage[final]{pdfpages}
\usepackage{comment}
\usepackage{tcolorbox}

\makeatletter
\AtBeginDocument{\let\LS@rot\@undefined}
\makeatother

\begin{document}
\title{Asymptotic versus mesoscopic spectral dimensions in networks and inhomogeneous structures}
%Dimensional Decoupling in Composite Networks
\author{Lorenzo Grimaldi}
\email{lorenzo.grimaldi@cref.it}
\affiliation{`Enrico Fermi' Research Center (CREF), Via Panisperna 89A, 00184 - Rome, Italy}
\affiliation{Dipartimento di Fisica dell'Università di Roma "Tor Vergata" e sezione INFN di Roma "Tor Vergata", Via della Ricerca Scientifica 1, 00133 - Rome, Italy}
\author{Pablo Villegas}
\email{pablo.villegas@cref.it}
\affiliation{`Enrico Fermi' Research Center (CREF), Via Panisperna 89A, 00184 - Rome, Italy}
\affiliation{Instituto Carlos I de F\'isica Te\'orica y Computacional, Univ. de Granada, E-18071, Granada, Spain.}
\author{Alessandro Vezzani}
\affiliation{Istituto dei Materiali per l'Elettronica ed il Magnetismo (IMEM-CNR), Parco Area delle Scienze, 37/A-43124 Parma, Italy}
\affiliation{Dipartimento di Scienze Matematiche, Fisiche e Informatiche,
Universit\`a degli Studi di Parma, Parco Area delle Scienze, 7/A 43124 Parma, Italy}
\affiliation{INFN, Gruppo Collegato di Parma, Parco Area delle Scienze 7/A, 43124 Parma, Italy}
\author{Raffaella Burioni}
\affiliation{Dipartimento di Scienze Matematiche, Fisiche e Informatiche,
Universit\`a degli Studi di Parma, Parco Area delle Scienze, 7/A 43124 Parma, Italy}
\affiliation{INFN, Gruppo Collegato di Parma, Parco Area delle Scienze 7/A, 43124 Parma, Italy}
\author{Davide Cassi}
\affiliation{Dipartimento di Scienze Matematiche, Fisiche e Informatiche,
Universit\`a degli Studi di Parma, Parco Area delle Scienze, 7/A 43124 Parma, Italy}
\affiliation{INFN, Gruppo Collegato di Parma, Parco Area delle Scienze 7/A, 43124 Parma, Italy}
\author{Andrea Gabrielli}
\affiliation{`Enrico Fermi' Research Center (CREF), Via Panisperna 89A, 00184 - Rome, Italy}
\affiliation{Dipartimento di Ingegneria Civile, Informatica e delle Tecnologie Aeronautiche, Universit\`a degli Studi “Roma Tre”, Via Vito Volterra 62, 00146 - Rome, Italy.}
          
\begin{abstract}
"Every object that biology studies is a system of systems." (François Jacob, 1974). Most networks feature intricate architectures originating from tinkering, a repetitive use of existing components where structures are not invented but reshaped. Linking the properties of primitive components to the emergent behavior of composite networks remains a key open challenge. Here, we show that modular compositions generically decouple the spectral and Fiedler dimensions, so that networks with identical thermodynamic exponents can exhibit parametrically different equilibration dynamics.
\end{abstract}

\maketitle

%\section{Introduction}
Nature often builds complexity by integration \cite{Jacob1977}. Across biological and artificial systems, new functions arise through the reuse and recombination of pre-existing modules \cite{Crisp2015,Pace2008,Grillner2016,Villegas2020,Fortuna2011}. This suggests viewing many real networks as composite systems assembled from distinct functional units. Understanding how internal module structure and inter-module organization combine to generate collective behavior remains a central challenge in complex systems.

In this context, the concept of dimension provides a quantitative bridge between topology, geometry, and dynamics, shaping diffusion, synchronization, and transport \cite{Mandelbrot,Orbach1986}. Among the various notions of dimension, the \emph{spectral dimension} $d_s$ (defined from the scaling of the Laplacian eigenvalue density) plays a central role \cite{AlexanderOrbach,Hattori1987,Cassi1992,Cassi1996, Burioni1996,spectral_dim_univ,RW_review,Villegas2025,Poggialini2025}. It governs asymptotic return probabilities, density-of-states singularities, and thermodynamic critical behavior, and is therefore often taken as a primary descriptor of network dynamics \cite{Burioni1996, spectral_dim_univ,RW_review}. 

Yet, the spectral dimension is defined in the thermodynamic limit. At finite but large size $N$, additional structural indicators become relevant. In particular, the smallest non-vanishing Laplacian eigenvalue—the Fiedler eigenvalue—sets the slowest relaxation time and controls equilibration and diffusion times \cite{Pecora1998,Barahona2002,Jiang2023}. Its scaling defines the Fiedler dimension $d_g$ \cite{Fiedler_eig}. In homogeneous networks, the spectral dimensions and Fiedler dimensions coincide \cite{Villegas2025,spectral_fiedler_instability}, ensuring that the bulk thermodynamic scaling and global equilibration are governed by the same effective geometry. In this case, the scaling of the density of states and the scaling of the equilibration time remain tightly linked.

Composite networks challenge this correspondence. Bundled networks—a paradigmatic class of network of networks—capture essential features of hierarchical assembly and can show anomalous diffusion \cite{Cassi1996_Bundled,Forte2013}. While their spectral properties are partly understood, how their slowest modes scale remains unclear \cite{spectral_dim_bund,RW_review}.

Such a separation has important physical implications. If the spectral dimension controls density-of-states scaling and critical behavior, the Fiedler dimension controls equilibration times, suggesting that composite architectures may violate single-parameter scaling: networks with identical thermodynamic exponents can exhibit parametrically different relaxation dynamics. This implies that bulk transport properties and terminal equilibration times need not be governed by the same geometric mechanism. In this sense, composite networks realize a geometry in which different effective dimensions control the thermodynamic limit and the approach to equilibrium.

Here, we show that in composite networks the Fiedler dimension generically decouples from the spectral dimension. In summary, we demonstrate that a single effective geometric dimension is insufficient to characterise dynamics in modular networks. Consequently, networks with identical spectral dimension can exhibit parametrically different equilibration times. Using diffusion-scale spectral analysis within the Laplacian Renormalization Group framework \cite{LRG,InfoCore,Modularity} together with explicit analytical calculations, we derive general expressions for both dimensions in bundled architectures. Our results demonstrate that bulk spectral properties and global relaxation are governed by distinct geometric mechanisms, revising the common assumption that a single effective dimension controls network dynamics.

\paragraph*{\textbf{Detecting scale-invariance in elementary and composite networks.}}
To probe diffusion scales, we analyze the Laplacian time-evolution operator $e^{-\tau \hat L}$, where $\hat L=\hat D-\hat A$ is the Laplacian operator for a connected undirected graph\cite{InfoCore}. Denoting its eigenvalues by $0=\lambda_0<\lambda_1\le\dots\le\lambda_{N-1}$, we introduce the Laplacian density matrix \cite{Domenico2016, LRG,Modularity},
\begin{equation}
\hat{\rho}(\tau) = \frac{e^{-\tau \hat{L}}}{Z(\tau)} = \frac{e^{-\tau \hat{L}}}{\sum_{i=0}^{N-1} e^{-\lambda_i \tau}}\,,
\label{RhoMat}
\end{equation}
which progressively suppresses high-frequency modes as $\tau$ increases, allowing for a canonical-ensemble description of heterogeneous networks \cite{LRG, InfoCore}. In particular, the diffusion time \( \tau \) is a scale parameter playing the role of an inverse temperature, where \( \hat{L} \) acts as a Hamiltonian, and \( Z(\tau) \) is the partition function. The associated entropy $S(\tau)=-\mathrm{Tr}[\hat\rho(\tau)\log\hat\rho(\tau)]$ defines the spectral susceptibility \cite{InfoCore}
\begin{equation}
C(\tau) \equiv -\frac{dS}{d \log \tau} = -\tau^2 \frac{d \langle \lambda \rangle_\tau}{d \tau},
\label{SHeat}
\end{equation}
which acts as a diffusion-scale heat capacity \cite{LRG,InfoCore}. Analogously to thermodynamic phase transitions, peaks in \( C(\tau) \) indicate structural transitions. Analyzing \( C(\tau) \) thus reveals the multi-scale organization of the network.

For large networks with power law spectral density $\rho(\lambda)\sim\lambda^{d_s/2-1}$ at small $\lambda$ \cite{AlexanderOrbach,Hattori1987,Burioni1996, spectral_dim_univ,RW_review}, one obtains $C(\tau)=d_s/2$ over the corresponding scaling regime \cite{Poggialini2025,Villegas2025}. 
Thus, the plateaus in $C(\tau)$ directly measure the spectral dimension $d_s$.

At finite size, however, the spectrum is discrete, and the smallest non-vanishing eigenvalue $\lambda_1\equiv\lambda_g$ controls the relaxation or equilibration time $\tau_\mathrm{eq}\sim\lambda_g^{-1}$. Since $C(\tau)$ vanishes for $\tau\gtrsim\lambda_g^{-1}$, the scaling $\lambda_1\sim N^{-2/d_g}$ defines a \emph{Fiedler or gap dimension} $d_g$ \cite{Fiedler_eig}. In homogeneous networks $d_g=d_s$ \cite{Villegas2025,spectral_fiedler_instability}, so that thermodynamic scaling and equilibration are governed by the same effective dimension. This equivalence reflects the fact that, in homogeneous geometries, the thermodynamic limit and the long-time limit lead to consistent scaling behaviors: density-of-states scaling and gap-controlled relaxation are governed by the same effective dimension. Here, we show that in composite architectures, this correspondence breaks down. This decoupling has direct consequences for finite-size response functions: the intermediate-time scaling is governed by $d_s$, while the terminal relaxation is controlled by $d_g$ [see Supplemental Material (SM) \cite{SI}, Secs.~S1--S2].

\paragraph*{\textbf{Regular fractal networks.}}

As a validation of the method, we consider two homogeneous fractal graphs: the Sierpinski gasket and the Sierpinski carpet (see Fig.~1 in SI1). In both cases, the heat capacity $C(\tau)$ exhibits oscillations associated with discrete self-similarity \cite{Oscill_gasket,Osscill_carpet,Oscill_fractals,Oscill_gasket2, GabrielliBook}, superimposed on a plateau whose value corresponds to $d_s/2$ \cite{Poggialini2025}. 

For the gasket, the plateau yields $d_s/2\simeq 0.684$, consistent with the exact value $d_s=\log 3/\log 5$ \cite{Hilfer1984,Rammal1984}. Independently, the scaling of the Fiedler eigenvalue with system size gives $d_g/2\simeq 0.683$, confirming $d_g=d_s$. Similarly, for the carpet, both the plateau analysis and the gap scaling yield consistent estimates of $d_s/2\simeq 0.9$ \cite{Barlow1990}. 

These results, as expected, confirm that in homogeneous fractal geometries the spectral and Fiedler dimensions coincide.

\paragraph*{\textbf{Dimension decoupling in composite networks.}}
We now consider bundled networks, constructed by attaching a copy of a generic graph (the \emph{fiber}) to each node of a base network \cite{Cassi1996_Bundled}. These architectures are intrinsically inhomogeneous and have long been known to exhibit anomalous transport properties \cite{Weiss1986,Cassi1996_Bundled,Boson_comb1,Boson_comb2,Boson_comb3,Comb_Bertacchi,comb_of_comb}.

An important property of bundled networks is that the spectral dimension coincides with that of the fiber, $d_s=d_{s,f}$ \cite{spectral_dim_bund,RW_review}, since in the thermodynamic limit, in which the sizes of the base and of the single fiber scale proportionally, the fraction of nodes belonging to the base vanishes compared to that of the fibers.

\begin{figure}[hbtp]
    \centering
    \includegraphics[width=\linewidth]{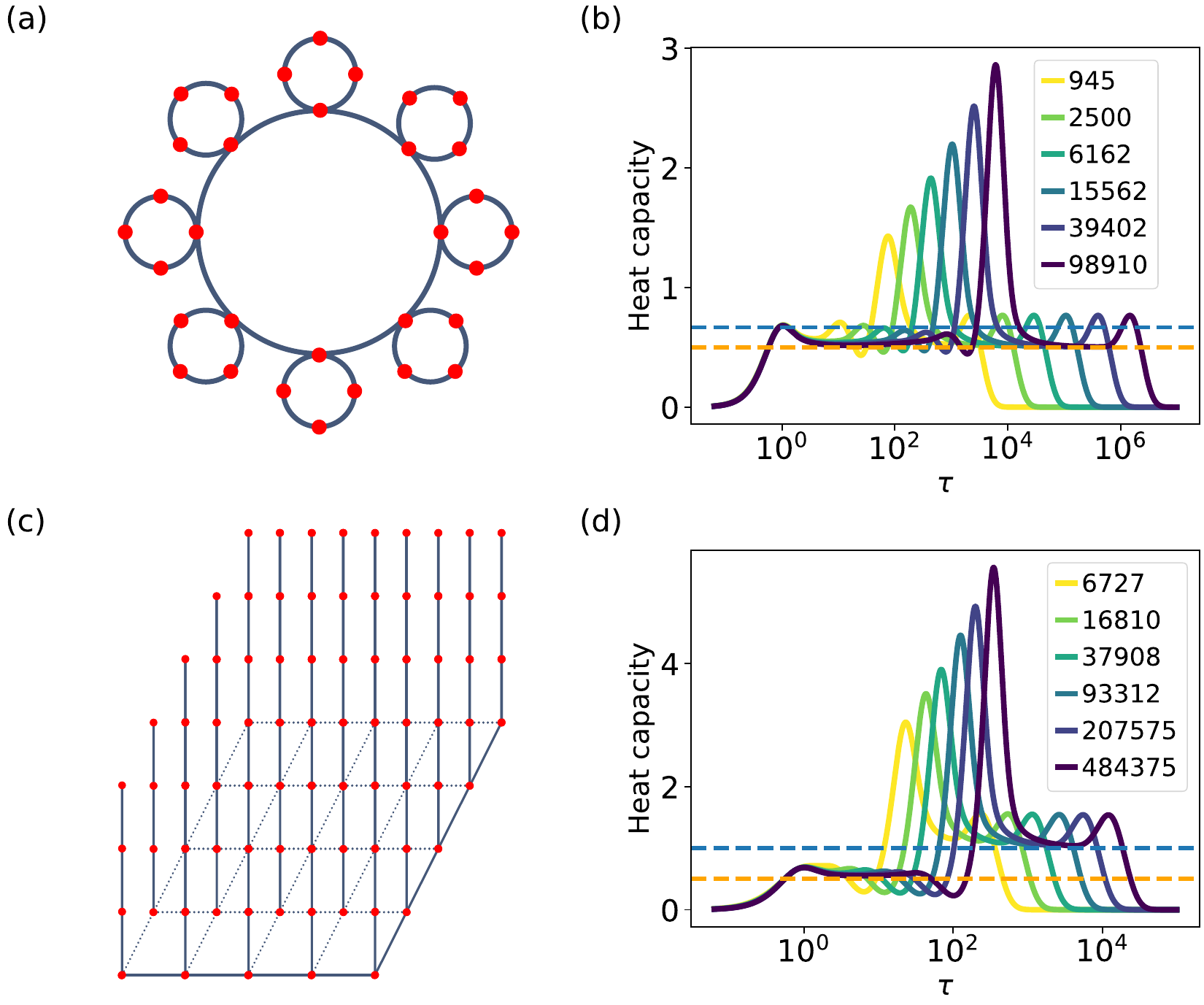}
    \caption{\textbf{(a)} Dirac comb (ring-of-rings) network. 
    \textbf{(b)} Heat capacity $C(\tau)$ versus diffusion time $\tau$ for the periodic comb. 
    The first and second plateaus correspond to the spectral dimensions of the fiber and the base, respectively. Since both are one-dimensional, $C=d_s/2=1/2$ (yellow dashed line). The blue dashed line indicates the Fiedler dimension predicted by Eq.~\eqref{dg_gen}. 
    \textbf{(c)} Dirac brush network (periodic boundary conditions omitted for clarity). 
    \textbf{(d)} Heat capacity $C(\tau)$ for the periodic brush. The first plateau reflects the spectral dimension of the one-dimensional fibers, while the second plateau corresponds to the two-dimensional base. Despite identical spectral dimensions, the two architectures exhibit different large-$\tau$ cutoffs, reflecting distinct Fiedler dimensions (blue dashed line).}
    \label{comb and brush}
\end{figure}

To illustrate the multi-scale structure, we consider two paradigmatic examples: the Dirac comb (ring of rings) and the Dirac brush (two-dimensional base with one-dimensional fibers) (see Fig.~\ref{comb and brush}). We set periodic boundary conditions to avoid border effects, thus enhancing numerical convergence. In both cases, we consider finite-size systems where the linear size $L$ of the base and of the fibers coincide.

In both cases, as reported in Figs.~\ref{comb and brush}(b) and (d), $C(\tau)$ exhibits multiple scaling regimes. The first plateau corresponds to the fiber dimension $d_{s,f}/2$ and dominates in the thermodynamic limit. For finite systems, this first plateau is followed by a global maximum that diverges with the system size. At larger diffusion scales, a second plateau emerges reflecting the base dimension, $C(\tau)=C_1=d_{s,b}/2$. Finally, the large-$\tau$ cutoff is set by the spectral gap $\lambda_g$, indicating that equilibration is controlled by the Fiedler scaling. These regimes are fingerprints of how the composite geometry reorganizes the low-energy spectrum and, in particular, the scaling of the spectral gap.

To analyze the low-energy spectrum, let $\hat L^b$ and $\hat L^f$ denote the Laplacians of base and fiber. Writing the eigenvalue equation for the global Laplacian operator in the basis diagonalizing $\hat L^b$ \cite{Boson_comb1,Boson_comb2,Boson_comb3}, one obtains
\begin{equation}
\sum_{j_2}\hat L^f_{i_2,j_2}\tilde\psi_{n_1,j_2}
+ \ell^b_{n_1}\delta_{i_2,0}\tilde\psi_{n_1,i_2}
= \lambda \tilde\psi_{n_1,i_2},
\end{equation}
where $\ell^b_{n_1}$ are $\hat L^b$ eigenvalues and $\tilde\psi_{n_1,j_2}$ are the eigenvectors of the global Laplacian already diagonalized in $\hat L^b$ (see SI2). This corresponds to a fiber with a localized impurity of strength $\ell^b_{n_1}$. Since the perturbation is confined to a single site, the bulk density of states remains governed by the fiber spectrum, yielding $d_s=d_{s,f}$.

The lowest eigenvalues arise from small $\ell^b_{n_1}$. A perturbative analysis (see SI2 and SI3) yields
\begin{equation}
\lambda_k \sim \frac{\ell^b_k}{N_f},
\label{lambda_pert}
\end{equation}
where $N_f$ is the number of nodes per fiber.

Since $\ell^b_1 \sim N_b^{-2/d_{g,b}}$, where $d_{g,b}$ is the base Fiedler dimension, and using $N_b \sim L^{d_{f,b}}, ~N_f \sim L^{d_{f,f}}, ~N \sim L^{d_{f,b}+d_{f,f}},$ with $d_{f,b}$ and $d_{f,f}$ being respectively base and fiber Hausdorff fractal dimensions, we obtain
\begin{equation}
\lambda_1 \sim 
N^{-\frac{d_{f,f}+2\frac{d_{f,b}}{d_{g,b}}}{d_{f,f}+d_{f,b}}},
\end{equation}
being the Fiedler dimension of the composite network,
\begin{equation}
d_g = 2\frac{d_{f,f}+d_{f,b}}{d_{f,f}+2\frac{d_{f,b}}{d_{g,b}}}.
\label{dg_gen}
\end{equation}

Equation~\eqref{dg_gen} shows that $d_g$ generically differs from both $d_{g,b}$ and the fiber spectral dimension $d_{s,f}$. This provides a composition rule
for the Fiedler dimension in modular architectures. Composite architectures, therefore, violate single-parameter geometric scaling: the thermodynamic limit and the approach to equilibrium probe different effective dimensions. Only when $d_{g,b}=2$ or when fibers remain finite does one recover $d_g=d_{g,b}$.

Importantly, Eq.~\eqref{dg_gen} depends only on the scaling exponents of the base and fiber components and is therefore independent of the microscopic details of the network architecture. Provided that the constituent networks possess a finite Hausdorff dimension, the result applies to controlled modular compositions beyond the specific examples considered here. The decoupling between $d_s$ and $d_g$ thus emerges as a generic geometric property of composite architectures rather than a feature of specific models.

\begin{figure}[h]
    \centering
    \includegraphics[width=\linewidth]{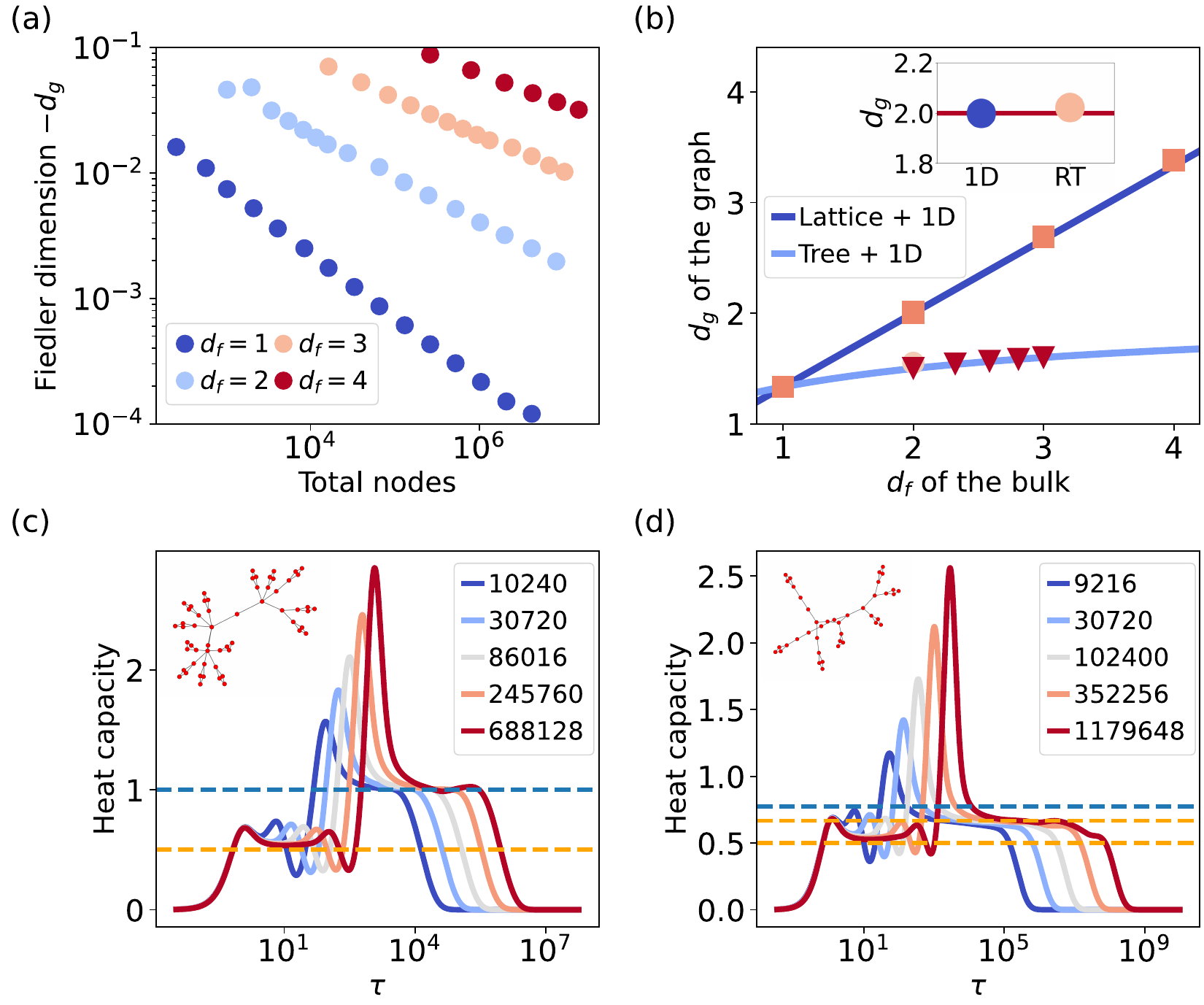}
    \caption{
    \textbf{(a)} Finite-size convergence of the Fiedler scaling: deviation of the local exponent from the analytical prediction of Eq.~\eqref{dg_gen} versus system size for lattice bases of varying dimension (see legend). 
    \textbf{(b)} Numerical validation of Eq.~\eqref{dg_gen} across composite families. Squares: $d$-dimensional lattice bases with one-dimensional ring fibers. Triangles: fractal-tree bases with ring fibers (see SI4 and SI5). The agreement confirms that the Fiedler dimension is fully determined by the composition rule. Inset: spectral dimension for BA networks with ring and random-tree fibers, confirming that $d_s$ is fiber-controlled. 
    \textbf{(c, d)} Heat capacity $C(\tau)$ for increasing system sizes in (c) BA–ring and (d) RT–ring architectures. The first plateau reflects the spectral dimension of the fibers, the second that of the base. The large-$\tau$ cutoff follows the Fiedler dimension (blue dashed lines).}
    \label{ba rt ls df}
\end{figure}

We validate Eq.~\eqref{dg_gen} across several composite families, including lattice–ring, tree–ring, and Barabási–Albert (BA) bases with ring or random-tree fibers (see Fig.~\ref{ba rt ls df}). In all cases, numerical scaling of the spectral gap agrees with the analytical prediction. Notably, for BA bases with effective $d_{g,b}=2$, the Fiedler dimension remains unchanged, consistent with Eq.~\eqref{dg_gen}. 

The physical implications of this decoupling are immediate in linear relaxation processes (see SI6). Within Rouse and generalized Gaussian network models \cite{Muller2003,Burioni2002409}, governed by $\dot x = -\hat L x$, each eigenmode relaxes as $e^{-\lambda_k t}$, so that the relaxation modulus reads $G(t)\propto \sum_k e^{-\lambda_k t}$.
In the thermodynamic limit this becomes $G(t)\sim \int \rho(\lambda)e^{-\lambda t} d\lambda$.
If $\rho(\lambda)\sim\lambda^{d_s/2-1}$, Laplace asymptotics give $G(t)\sim t^{-d_s/2}$ at intermediate times, showing that the spectral dimension governs bulk decay
(see Fig.\ref{fig:relax}).
For finite systems, relaxation is cut off at $t\sim\lambda_g^{-1}$, yielding $\tau_{\mathrm{eq}}\sim\lambda_g^{-1}\sim N^{2/d_g}$. The role of the Fiedler dimension in the equilibration times is evidenced in Fig.\ref{fig:relax}(b), where the time is rescaled by $\tau$, using $d_g$. A single scaling function is obtained for different sizes. This scaling function is driven by the lowest eigenvalues in Eq. \eqref{lambda_pert} and hence by the spectral dimension of the base $d_{s,b}$ (see SI6).

Finite composite networks, therefore, separate intermediate-time scaling (controlled by $d_s$) from equilibration times (controlled by $d_g$ and by $d_{s,b}$). This separation produces a two-regime relaxation behavior, with bulk rheology governed by the spectral dimension and terminal equilibration controlled by the spectral gap and by the lowest energy states, which in this specific case are confined on the base. As illustrated in Fig.~\ref{fig:relax}, brush and comb architectures share identical low-$\lambda$ scaling and intermediate decay,
since they have the same spectrum and spectral dimension in the thermodynamic limit. Yet, they exhibit parametrically different equilibration times due to distinct spectral-gap scaling, and different finite scaling functions due to different base topology.
This provides a minimal spectral mechanism for the anomalous or multimodal relaxation spectra (given by the Fourier transform of $G(t)$) observed in complex polymer architectures \cite{Kapnistos2008, Ge2023}. In this interpretation, the exponent of the relaxation modulus directly reflects the spectral dimension of the effective network architecture that governs stress propagation. Moreover, the spectral framework naturally leads to the double-scaling relaxation behavior reported in polymer rheology studies \cite{Kapnistos2009,Daniels2001,Inkson2006}.

\begin{figure}
\centering
\includegraphics[width=0.494\linewidth]{RelaxationModulus_edited.pdf}
\includegraphics[width=0.494\linewidth]{RelaxationModulusRescaled_edited.pdf}
\caption{\label{fig:relax}
$G(t)\propto \langle e^{-\lambda t}\rangle$ computed from the Laplacian spectrum of brush (blue) and comb (orange) architectures.
\textbf{(a)} In the thermodynamic limit, both systems exhibit the same intermediate-time decay $G(t)\sim t^{-d_s/2}$, reflecting their identical spectral dimension $d_s$. At long times, however, finite-size effects lead to distinct equilibration behaviors.
\textbf{(b)} Rescaling time by the equilibration time $\tau_{\mathrm{eq}}\sim N^{2/d_g}$ produces a collapse onto distinct scaling functions, demonstrating that equilibration is controlled by the Fiedler dimension $d_g$. The different scaling functions further reflect the influence of the base structure through its low-energy spectrum.
}
\end{figure}

An analogous separation arises in vibrational thermodynamics. In elastic network models, Laplacian eigenvalues determine normal-mode frequencies and thus the Debye heat capacity (see SI7). While the low-temperature scaling $C_V \sim T^{d_s}$ is governed by the spectral dimension in the thermodynamic limit, finite-size crossover occurs at a temperature set by the spectral gap, $T_c \sim N^{-1/d_g}$. Composite architectures, therefore, decouple thermodynamic scaling from finite-size crossover: the bulk temperature dependence reflects $d_s$, whereas the finite-size cutoff is controlled by $d_g$ and by the spectrum of the base network (see SI2 and SI3). This demonstrates that thermodynamic scaling and dynamical equilibration probe distinct effective geometries in composite networks.

\paragraph*{\textbf{Outlook.}} Natural and artificial complex systems often emerge not through top-down design but through a process of tinkering—the gradual recombination of pre-existing modules into new functional architectures \cite{Jacob1977}. Such compositional growth does more than simply increase size: it reshapes geometry itself \cite{Fortuna2011, Villegas2020}.

We have shown that dimensionality in composite networks is not a single invariant quantity. While the spectral dimension governs thermodynamic scaling and density-of-states behavior, the Fiedler dimension controls equilibration and spectral-gap scaling. In homogeneous geometries, these quantities coincide, yielding an effective single-parameter description. In modular architectures, however, they generically decouple. Dimensionality thus emerges as a relational property of compositional architectures rather than an intrinsic attribute
of individual components.

This anomalous behavior reflects a subtle interplay between the dimensions of the constituent elements that goes beyond a simple sum or average. As a consequence, the thermodynamic limit and the long-time limit need not commute: two systems can share identical bulk exponents while exhibiting parametrically different relaxation times. Finally, the composition rule for $d_g$ suggests a natural extension to hierarchical architectures, where successive layers of modular assembly may generate a hierarchy of effective dimensions controlling different relaxation scales.

In more general networked systems, compositional organization may involve several structural scales rather than a single base--fiber decomposition. In that case, the separation between density-of-states scaling and spectral-gap scaling need not be captured by a single pair of dimensions, but may instead give rise to crossovers between different effective geometries, or to a hierarchy of dimensions associated with successive levels of modular assembly. The present construction should therefore be regarded as a minimal controlled setting in which this dimensional decoupling can be isolated analytically.

\begin{acknowledgments}
\paragraph*{\textbf{Acknowledgments--}}P.V. acknowledges the Spanish Ministry of Research and Innovation and Agencia Estatal de Investigación (AEI), MICIN/AEI/10.13039/501100011033, for financial support through Project PID2023-149174NB-I00, funded also by European Regional Development Funds, and  Ref. PID2020-113681GB-I00.
\end{acknowledgments}

%\bibliography{sample}

\clearpage
\includepdf[pages={1}]{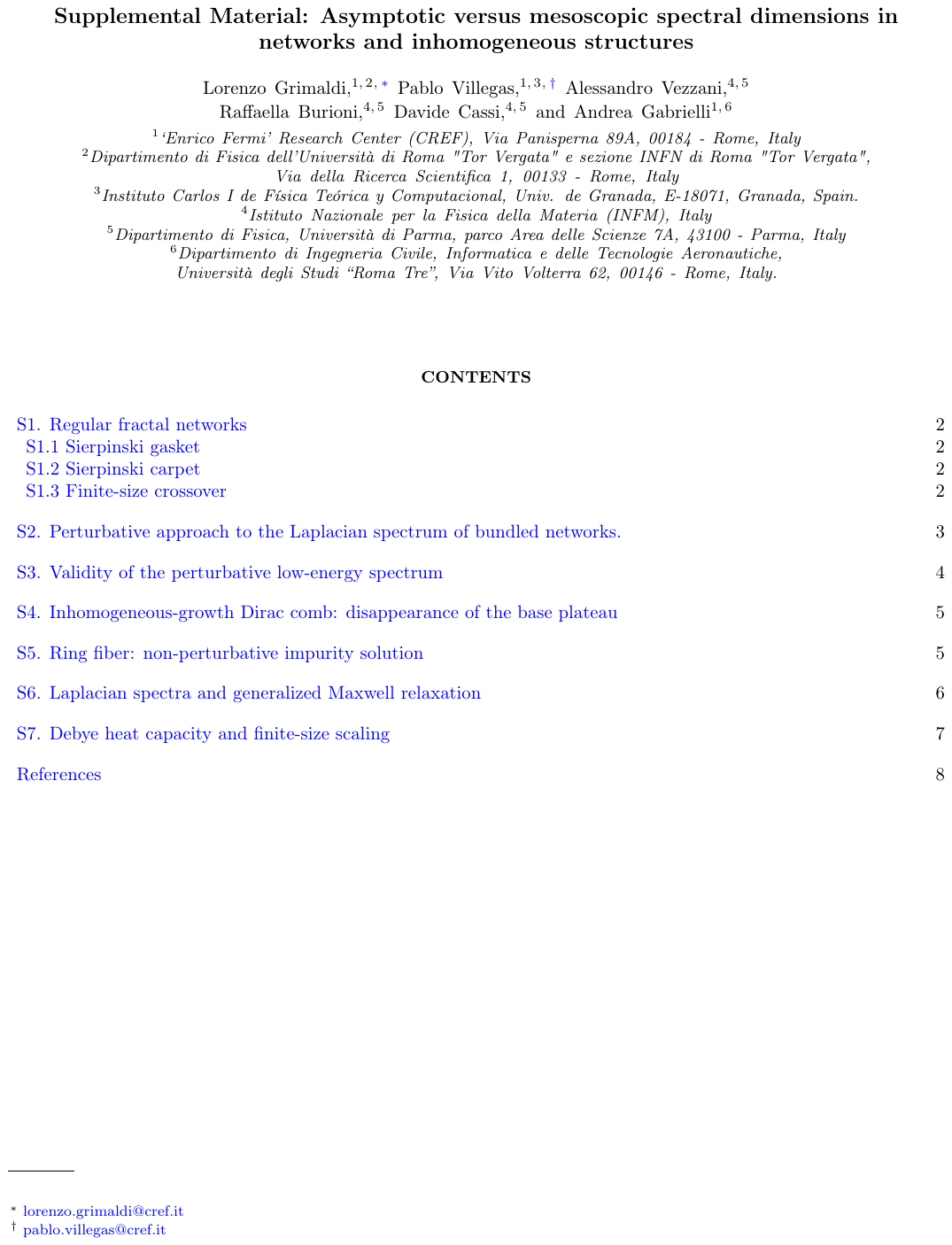}
\clearpage
\includepdf[pages={2}]{1-SupInf.pdf}
\clearpage
\includepdf[pages={3}]{1-SupInf.pdf}
\clearpage
\includepdf[pages={4}]{1-SupInf.pdf}
\clearpage
\includepdf[pages={5}]{1-SupInf.pdf}
\clearpage
\includepdf[pages={6}]{1-SupInf.pdf}
\clearpage
\includepdf[pages={7}]{1-SupInf.pdf}
\clearpage
\includepdf[pages={8}]{1-SupInf.pdf}
\clearpage
\end{document}